# A novel interface for adversarial trivia question-writing

Jason Liu
Montgomery Blair High School

*Mentors:*
Jordan Boyd-Graber
University of Maryland

Saptarashmi Bandyopadhyay
University of Maryland

# Abstract

A critical component when developing question-answering AIs is an adversarial dataset that challenges models to adapt to the complex syntax and reasoning underlying our natural language. Present techniques for procedurally generating adversarial texts are not robust enough for training on complex tasks such as answering multi-sentence trivia questions. We instead turn to human-generated data by introducing an interface for collecting adversarial human-written trivia questions. Our interface is aimed towards question writers and players of Quiz Bowl, a buzzer-based trivia competition where paragraph-long questions consist of a sequence of clues of decreasing difficulty. To incentivize usage, a suite of machine learning-based tools in our interface assist humans in writing questions that are more challenging to answer for Quiz Bowl players and computers alike. Not only does our interface gather training data for the groundbreaking Quiz Bowl AI project QANTA, but it is also a proof-of-concept of future adversarial data collection for question-answering systems. The results of performance-testing our interface with ten originally-composed questions indicate that, despite some flaws, our interface's novel question-writing features as well as its real-time exposure of useful responses from our machine models could facilitate and enhance the collection of adversarial questions.

The code for our interface is available at: https://github.com/Zefan-Cai/QAML

# 1. Quiz Bowl and machine learning

Quiz Bowl is a team competition of trivia question-answering. The central component of Quiz Bowl is the "tossup" question, which is read while players on each team compete to be the first to interrupt with the correct answer using a buzzer system (NAQT c). These "questions" are not phrased as questions in the grammatical sense (e.g. "Who was the first president of the United States?"), but rather as a sequence of statements that act as clues unique to the answer (e.g. "This man was the first president of the United States."). The difficulty of the clues is intended to decrease as the question progresses—a concept known as "pyramidal" question-writing in Quiz Bowl (NAQT a). For instance, consider the following original Quiz Bowl tossup that we composed for demonstration purposes in this study:

> **Question:** This man was inspired by the work of the composer Toru Takemitsu while staying in Japan, and conducted the premiere of Takemitsu's *Dorian Horizon*. This man formed a friendship with the Mexican composer Carlos Chávez to whom his second symphony, the *Short Symphony*, is dedicated. This man also wrote the music for the opera *The Tender Land* and later composed the orchestral work *Connotations*. This composer's third and final symphony was premiered by Serge Koussevitzky, and its final movement forms the basis for his *Fanfare for the Common Man*. For ten points, name this American composer known for works such as *Rodeo* and *Appalachian Spring*.
>
> **Answer:** Aaron Copland

In this example, clues begin with relatively obscure details such as Copland's connections to Takemitsu and Chávez, and end on more well-known facts like *Appalachian Spring*. For players to buzz in earlier in the tossup, they must acquire a more in-depth understanding of the question's topics. The same is true for machines.

A computer's ability to buzz early and accurately on Quiz Bowl questions not only indicates how well it can synthesize and interrelate facts drawn from large sources of information (e.g. Wikipedia dumps, Quiz Bowl question datasets), but also gauges its capacity to understand human language and reasoning (Rodriguez et al. 2021), the challenges of which are elaborated in **Adversarial question-writing**. Thus, it is Quiz Bowl's pyramidal question-writing and competitive buzzing that makes it the useful focus of many question-answering studies in machine learning. Such research includes the development of QA models for playing Quiz Bowl (Iyyer et al. 2014; Yamada et al. 2018; Rodriguez et al. 2021), as well as human-in-the-loop interfaces for collecting adversarial questions that are used to improve the robustness of these models (Feng and Boyd-Graber 2019; Wallace et al. 2019; Eisenschlos et al. 2021).

## 1.1 Adversarial question-writing

A critical factor in Quiz Bowl models is their ability to understand and adapt to natural language, including the unique challenges that arise from Quiz Bowl question-writing. We intend to achieve this by gathering an adversarial dataset that targets common Quiz Bowl AI pitfalls, which Wallace et al. 2019 divide into two adversarial categories: distraction and reasoning.

Distraction involves clues intentionally worded to confuse the computer and is the more common of the two categories; 71% of adversarially-authored Quiz Bowl questions collected by Wallace et al. contain distractions. Distractive modifications can be syntactic-level, such as paraphrasing "This man was inspired by the work of Toru Takemitsu" from the example tossup as "The work of Toru Takemitsu inspired this man", or semantic-level, such as substituting "its final movement" for the meaningfully equivalent "its finale". Paraphrasing is recognized as an adversarial example, since neural networks have been shown to be consistently susceptible to misclassifying paraphrased forms of examples from their training datasets (Goodfellow et al. 2015). The introduction of novel clues that can not be learned from the dataset also proves troublesome for computers. Neural models in particular could be tricked into giving the incorrect answer when faced with novel clues (Wallace et al. 2019).

Reasoning encompasses the computer's ability to logically and factually reach an answer using the knowledge it has already gleaned from its training. A common form of this is multi-step reasoning, which is found in 25% of adversarial questions and involves using connections between multiple entities to arrive at the final answer (Wallace et al. 2019). Consider the following from the example tossup: "This man also wrote the music for the opera *The Tender Land*". We can rewrite this clue to incorporate multi-step reasoning, making the question harder for the machine to answer as follows: "James Agee's book on Southern sharecroppers during the Great Depression was an inspiration for a 1954 opera that this man composed." No longer is the opera's name explicitly mentioned; rather, the computer must first determine that the book referenced is *Let Us Now Praise Famous Men*, then associate the book with *The Tender Land*, before realizing that the question asks for the opera's composer and arriving at the final answer of Aaron Copland. Achieving multi-step reasoning in computers is still considered a complex challenge and is the subject of recent work on machine reading comprehension over text paragraphs of similar length to Quiz Bowl tossups (Shen et al. 2017; Liu et al. 2021). A related form of reasoning is multi-hop reasoning, defined to be reasoning requiring information from different parts of the question to arrive at the answer (Yang et al. 2018). Thus, multi-hop reasoning is inherent to most Quiz Bowl tossups due to their sequential structure of separate clues.

Wallace et al. identify human-in-the-loop question-writing to be the most viable technique for the generation of adversarial questions. While machine generation of minor textual perturbations such as typos and character replacements has been demonstrated (Wallace et al. 2019), broader and more complex tasks such as generating adversarial multi-hop reasoning examples and paraphrasing remain open challenges. HotpotQA, one of the leading datasets for question-answering with multi-hop reasoning, still required manual Wikipedia page curation and crowdsourcing for its data collection (Yang et al. 2018). Furthermore, most adversarial example generators are limited to sentences and do not generalize well to multi-sentence Quiz Bowl tossups. Syntactically controlled paraphrase networks, for example, are restricted to "purely syntactic modifications" (Iyyer et al. 2018) and produce low rates of valid paraphrases for Quiz Bowl questions (Wallace et al. 2019). Thus, we select a human-in-the-loop approach assisted by helpful feedback from exposed machine learning models (e.g. where the machine buzzes in the question, the machine's top answer guesses, clues the machine considers important for its buzz), which Wallace et al. argue makes the process model-driven while benefiting from human creativity. Our interface addresses and promotes the various forms of adversarial question-writing among users by highlighting key "giveaway" phrases that could be rewritten to increase question difficulty as well as by recommending underrepresented topics to encourage inclusion of novel clues and to improve dataset diversity.

## 1.2 Machine learning models

The interface developed in this study exposes key machine learning-based features to facilitate human-in-the-loop adversarial question-writing. These models are tried and tested in past research for developing QA models in Quiz Bowl. They are also trained on a common Quiz Bowl question dataset, known as QANTA (Question Answering is Not a Trivial Activity), consisting of over 100,000 human-authored English questions from the Quiz Bowl community (Rodriguez et al. 2021). The original QANTA Quiz Bowl AI was a recursive neural network model (Iyyer et al. 2014), though the QANTA dataset has since been used to develop a variety of other Quiz Bowl models, including information retrieval (IR) and Bidirectional Encoder Representations from Transformers (BERT) models (Rodriguez et al. 2021). For the purposes of our interface, we use IR models to build our machine buzzer, guesser, and pronunciation difficulty modules, and BERT-based models for classifying question difficulty and recommending underrepresented topics.

## 2. Developing the interface

The main objectives during our interface development were consolidating and improving existing adversarial Quiz Bowl question-writing interfaces. Namely, these are the question-writing-oriented Trick Me interface by Eric Wallace and the "gamified" Fool Me Twice interface by Julian Eisenschlos. The former's question-writing UI became the basis for our interface's modular workspace design, while the latter's point-based reward system was incorporated into our interface's optional "game mode". Furthermore, widgets that allow the user to interact with our machine learning modules and novel organizational features that support an efficient question-writing workflow were implemented on the frontend to incentivize usage of our interface.

Our interface's server was built with Flask in order to facilitate backend implementation of Python's machine learning libraries and features. The frontend was built using the Vue.js framework for component modularity, which encourages users to propose their own widget components to continually improve our interface.

## 2.1 Merging Trick Me and Fool Me Twice

A number of AI-based features were carried over from Wallace's Trick Me, including top answer guesses from the machine and highlighting for evidence the machine finds most important for buzzing. Though Wallace et al. identify "novel clues" and "unfamiliar data" among the deficiencies of both neural and IR models when playing Quiz Bowl, our interface was the first to present a widget that aids users in generating unfamiliar clues through exposure of underrepresentation data. This involved displaying question genre and subgenre distributions, recommending related underrepresented countries in real time, and rewarding users for submitting questions that are different from existing ones.

Our interface's gamification of question-writing adapted many functionalities that were tried and tested in Fool Me Twice. This included rewarding users with points for writing questions that the machine considered more adversarial and diverse. Another feature transferred from Fool Me Twice into our interface's game mode was question-writing under a time constraint, which Eisenschlos et al. believe improves user enjoyment.

## 2.2 Incentivizing human interaction

Our interface was developed with a mutually beneficial relationship in mind: the user is intended to use our interface's organizational and data-rich features to more effectively write questions that are challenging for both Quiz Bowl-playing AIs and humans.

To achieve this goal, our interface introduced novel features aimed at encouraging usage from professional Quiz Bowl question writers in particular. User flexibility and customization was emphasized in the design: users could create and organize modular workspaces that facilitate an efficient workflow. Each workspace was implemented with a question-writing space and toggleable widgets, which expose helpful data on the fly from our machine learning modules.

Frontend widgets addressed Quiz Bowl-specific problems. Similar to the Trick Me interface, our interface's buzzer widget was built to not only indicate where in the current user-written question the machine has buzzed, but also to identify sentence and word-level evidence that were "giveaway" clues for the machine's buzz. Users can then target these clues for replacement or modification using any of the techniques discussed in **Adversarial question-writing**. The novel pronunciation widget exposed a model trained to predict a word's pronunciation difficulty; this widget highlighted for which words the author would likely need to include an in-text pronunciation guide, an important Quiz Bowl standard that facilitates question reading in tournament play. Furthermore, a novel similar questions widget was implemented to improve question diversity. This widget found which questions in the QANTA Quiz Bowl database were most similar to the user's current text, allowing the user to adjust clue topics and phrasing to make their question more unique. Post-hoc features include a question difficulty model, which classifies user-written questions as either "high school" or "college"-level difficulty.

IR models using tf-idf were selected for widgets depending on faster response times, including machine guessing and buzzing as well as pronunciation highlighting, since they were simpler to implement and significantly smaller in size than BERT-based models. IR was also applied to widgets like similar questions that required comparison over a large dataset of documents. BERT-based models were used for widgets such as question difficulty and underrepresentation where greater accuracy was valued.

## 3. Testing with original questions

### 3.1 Question-writing

Ten original tossup questions with answers were composed by the author to performance-test the interface, to observe highlighting and machine response to the text, and to collect incremental edit history data from the server. Each tossup was written according to National Academic Quiz Tournaments (NAQT) rules and standards (NAQT a; NAQT c). Question genres were chosen according to the NAQT high school subject distribution scaled to ten questions, resulting in one fine arts, one geography, two literature, three science, and three history questions (NAQT b). After each question was manually typed into our interface, edit histories were saved in data dumps by the server, along with other exposed data including buzzes, machine guesses, and pronunciation and country underrepresentation highlighting.

## 3.2 Questions with interface highlighting

**Table 1.** Table of ten questions with interface highlighting. Phrases considered important influences on the machine's guess are highlighted in green, words considered difficult to pronounce by the machine are highlighted in cyan, underrepresented countries are highlighted in purple, and the character corresponding to the machine's final buzzer position is followed by "**[buzz]**". If the buzzer position regressed during the input process, the first correct buzz is denoted with "**[buzz]**", and the first incorrect buzz (if it exists) is denoted with "**[buzz]** (*incorrect answer*)". Note that these last two colors are not included in the interface.

1. **Question [Fine Arts]:** This man was inspired by the work of the composer Toru Takemitsu while staying in Japan, and conducted the premiere of Takemitsu's *Dorian Horizon*. This man formed a friendship with the Mexican composer Carlos Chávez to whom his second symphony, the *Short Symphony*, is dedicated. This man also wrote the music for the opera *The Tender Land* and later composed the orchestral work *Connotations*. This composer's third and final symphony was premiered by Serge Koussevitzky, and its final movement forms the basis for his *Fanfare for the Common Man*. For ten points, name this American composer known for works such as *Rodeo* and *Appalachian Spring*.**[buzz]**
**Answer:** Aaron Copland

2. **Question [History]:** This man noted the common physical characteristics of the Germanic peoples in his book *Germania*. He was also the son-in-law of Agricola and wrote the primary biographical source of the general and his conquest of Britain. In his final work, following a passage on the Great Fire of Rome, this historian wrote of Pontius Pilate and the execution of a man named "Christus". For ten points, name this Roman historian who authored the *Histories* and the *Annals*.**[buzz]**
**Answer:** Tacitus

3. **Question [Literature]:** In one episode in this work, the main character and his valet are received by Jesuits in Paraguay. Another scene sees the main character saved by Jacques the Anabaptist, who later drowns in Lisbon before the city is hit by an earthquake. Other characters in this work include the Manichaeist Martin, the chambermaid Paquette, and the Baron of Thunder-ten-Tronckh. In**[buzz]** a musical work based on this book, one character sings that "one finds this is the best of all possible worlds". That operetta is by Leonard Bernstein. For ten points, name this novella featuring Cacambo, Dr. Pangloss, Cunégonde, and the title character, a satire by Voltaire.
**Answer:** *Candide*

4. **Question [Science]:** According to Varignon's theorem, this quantity can be algebraically summed when applied at a single point. The cross product of dipole moment and electric field, this quantity is considered a pseudovector in three dimensions.**[buzz]** When integrating this quantity with respect to angular position, the result is mechanical work. This quantity is denoted by the letter tau and is equivalent to the moment of inertia times the angular acceleration. For ten points, name this rotational equivalent of force.**[buzz]**
**Answer:** torque

5. **Question [History]:** This man's brother-in-law was sent to investigate the Anahuac Disturbances. That relative, Martín Perfecto de Cos, surrendered at the Siege of Béxar. This man led an uprising during the Casa Mata Plan Revolution, leading to the fall of the emperor Agustín de Iturbide. As he rose to power, this man defeated the Barradas Expedition and formalized the "Siete Leyes". He later fought in the Pastry War when he famously lost and buried his leg with full military honors. For ten points, name this Mexican president and general**[buzz]** whose centralist rule**[buzz]** saw territorial losses during the Texas Revolution and the Mexican-American War.
**Answer:** Antonio López de Santa Anna

6. **Question [Science]:** The condensation of these compounds was researched by Alexander Borodin. It's not fertilizer, but these compounds were studied by Justus von Liebig, who coined the name for these molecules. Both Fehling's solution**[buzz]** and Tollen's reagent test can be used to distinguish these compounds from ketones. These compounds are commonly formed through the oxidation of alcohols like methanol. For ten points, name these compounds containing the formyl group "-CHO".**[buzz]**
**Answer:** aldehydes

7. **Question [History]:** This present-day country was home to the Mossi Kingdoms prior to French colonization. The Agacher Strip War was fought between this country and its northern neighbor over a border dispute. During a 2014 uprising in this country, its president Blaise Compaoré was pressured to resign. Another[buzz] president of this country seized power from Jean-Baptiste Ouédraogo and planted millions of trees. That president, Thomas Sankara, renamed the country from Upper Volta to its present name. For ten points, name this West African country whose capital is Ouagadougou.[buzz]
**Answer:** Burkina Faso

8. **Question [Literature]:** In the opening to one of this man's novels, a woman recognizes the Sinfonietta by Leoš Janáček while riding a taxi. Another character in that novel is tasked with rewriting the story *Air Chrysalis*. In the opening chapter of a different novel by this man, the main character listens to a Rossini overture while making spaghetti. That character later meets a war veteran from the Manchurian campaign and sits at the bottom of a well while searching for his missing cat. For ten points, name this Japanese author of *The Wind-Up Bird Chronicle*[buzz] and *1Q84*.[buzz]
**Answer:** Haruki Murakami

9. **Question [Science]:** One part of these structures is named for Janus and Epimethus, while another is associated with Methone. A theor[buzz] (*density waves*)y for these structures' formation was proposed by Édouard Roche, who also names a section between these structures. The existence of gaps in these structures[buzz] was suggested by Pierre-Simon Laplace. These gaps include those named after Colombo and Maxwell, and the largest is the Cassini Division. For ten points, name these structures orbiting the sixth planet from the Sun.[buzz]
**Answer:** rings of Saturn

10. **Question [Geography]:** A ruler in this city moved his residence to Yıldız Palace in fear of a seaside attack.[buzz] (*New York City*) It's not Rome, but this city is known for its seven hills, on which[buzz] stand structures such as the Nuruosmaniye Mosque and the Fatih Mosque. The architect Sedefkar[buzz] Mehmed Agha based the design of this city's Blue Mosque on an earlier Byzantine church in this city. Not far from there is the Topkapı Palace, which overlooks both the Golden Horn and the Bosporus. For ten points, name this metropolis between Europe and Asia, the largest city in Turkey.
**Answer:** Istanbul

## 3.3 Results

As seen in Figure 1 below, all widgets in the interface were capable of responding to user input with data from the machine learning models.

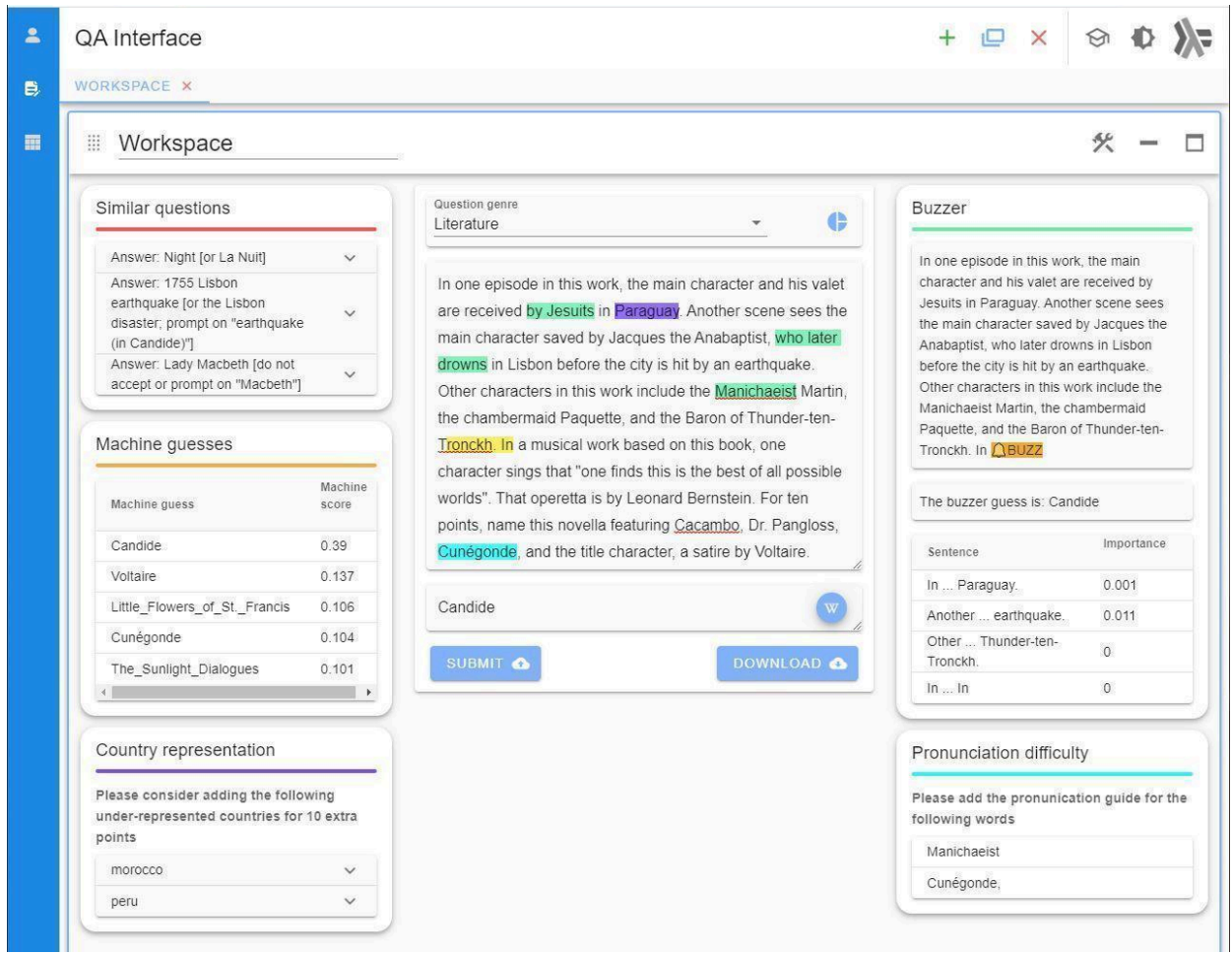

**Figure 1.** Screenshot of the interface after the entirety of Question 3 ("Candide") has been typed into the central input box by the user. Note not only the highlighting as seen in Table 1, but also the data provided in the widgets on the left (Similar questions, Machine guesses, Country Representation) and the right (Buzzer, Pronunciation difficulty) that can help Quiz Bowl writers improve their writing and provide more adversarial questions.

The buzzer module in particular was capable of buzzing with the correct answer by the end of every tossup question, but it was "tricked" into buzzing incorrectly after the first sentences of Question 9 ("rings of Saturn") and Question 10 ("Istanbul"). However, in most cases, the initial buzzer position noticeably "regressed" while the questions were being typed: the machine would correctly buzz on a clue that had just been enter, but the subsequent typing of more clues after the initial buzzer position prompted the server to reanalyze the question and buzz at a later position (Table 1). This behavior is troublesome, since users would expect the machine to consistently buzz after the same clue if no text before it has changed.

Our interface was capable of highlighting important phrases and difficult-to-pronounce words, but failed to consistently and accurately identify underrepresented countries. For example, the computer detected the inclusion of the Middle Eastern country "Oman" in Question 2 ("Tacitus"), but the actual word used was "R**oman**" (Table 1). In total, there was one true positive ("Paraguay" in Question 3) and three false positives ("Oman" in Questions 2 and 8, and "Mali" in Question 5).

The post-hoc question analysis during question submission to the database included passing each tossup through the BERT-based question difficulty classifier. The classifier, which was trained and tested to have a 79.95% accuracy rate in distinguishing between high school and college-level questions in the QANTA dataset, classified all ten of the questions as "high school"-level difficulty.

The server successfully captured question edit history in 10-20 second intervals. The data dumps collected by our interface yield time-based, granular text data that could assist the optimization of machine components such as the buzzer.

## 4. Discussion

Based on the results from testing our ten sample questions, we identified three troublesome aspects of our interface as well as potential solutions. In general, however, we believe our interface successfully met its goal of assisting humans in writing trivia questions that are more difficult for machines.

The regressing of the buzzer position was the result of a design choice when building our interface's buzzer module. The buzzer position was not "locked" in place unless the top machine guess—found through information retrieval on Wikipedia pages—both exceeded a certain confidence threshold as well as matched the user-provided answer. The sample questions revealed how such a design can lead to confusion on the user's side, especially since changes in buzzer position were not explained on the front-end. While the confidence threshold for buzzing can be adjusted given more trial-and-error testing, determining whether the user's inputted answer and the answer generated by the interface buzzer are equivalent is more nuanced. A more sophisticated entity linking algorithm is needed to match answers which are not visually similar but refer to the same entity. Even though the user-provided answers in the sample questions and their corresponding machine guesses were exactly the same, it is plausible, in Question 9 for example, that the user simply enters "Saturn's rings", which the current buzzer can not recognize as equivalent to its guess "Rings of Saturn", derived from the Wikipedia page title. A proposed fix to this issue, using Wikipedia redirect pages that connect similar entities like "Saturn's rings" to "Rings of Saturn", is already in testing on a development branch of our interface.

Our results exposed a more rudimentary flaw in the text-parsing method employed by the underrepresentation module. The current character-based analysis proved too granular and naive for identifying and highlighting underrepresented countries in the questions. Fixing this oversight primarily requires shifting from character-based to word-based search of key terms.

The final module that needed scrutiny was the post-hoc question difficulty classifier. Although the module classified all sample questions as "high school"-level difficulty, each question was purposely composed with college-level clues. The machine buzzer—trained on both high school and college-level questions—only buzzed correctly before the final "giveaway" clue in four of the questions (Table 1), suggesting at the very least that what the classifier considers to be in the "easier" class of questions does not correspond well with the capabilities of the machine buzzer. While the inclusion of college-level clues does not guarantee a college-level question, it is unlikely that all ten of these tossups would be categorized as high school-level difficulty in a real-world Quiz Bowl tournament given the challenge they posed to the buzzer. In this case, a more granular approach to implementing the BERT-based classifier, such as training it to identify the difficulty of a single sentence or clue rather than a whole question, could yield better results.

Compared to the interfaces developed by Wallace et al. 2019 and Rodriguez et al. 2021, we find that our interface features novel and functional modules. Despite the flawed implementation of the three modules discussed, our interface still exhibited unique benefits to question writers by exposing helpful machine data, such as pronunciation difficulty, that was not provided by previous interfaces. Suggestion of underrepresented countries coupled with identification of similar questions encourage improvement of question diversity; buzzer position and machine guesses allow assessment of question difficulty; and highlighting of key phrases "important" to the computer reveal where best to apply adversarial writing techniques to better challenge AIs.

## 5. Future work

By demonstrating a functional trivia-writing interface fitted with novel features, we show that a mutually beneficial future for both Quiz Bowl question-writing and question-answering AIs could be found in adversarial human-computer interaction. However, due to the paucity of both sample questions and participants for this study, analysis of our interface must continue over larger datasets and userbases to better determine its impact on the Quiz Bowl and NLP communities. We intend to launch our interface to the public in the coming months and begin formal user-testing in the same vein as Wallace et al. 2019 and Rodrigez et al. 2021, which involves collecting adversarial user-written questions through our interface, and evaluating their effect when training the QANTA Quiz Bowl AI. Further user-based research can also explore the effectiveness of our interface's game mode, and whether the time constraints benefit the user experience as well as increase the quantity of questions collected.

Overall, our interface proves a promising foundation for future work involving adversarial interfaces that are mutually beneficial to the Quiz Bowl community and natural language processing researchers. With the successful testing of novel features, we hope our interface will serve to improve the diversity and challenging nature of trivia questions for humans and AIs alike.

# 6. Acknowledgements

The author would like to acknowledge the mentorship of Dr. Jordan Boyd-Graber and Mr. Saptarashmi Bandyopadhyay, as well as the invaluable contributions from the following members of the interface development team:

Raj Shah
Georgia Institute of Technology

Atith Gandhi
Birla Institute of Technology & Science, Pilani

Zefan Cai
Peking University

Damian Rene
The Gunston School

Answering," *Transactions of the Association for Computational Linguistics*, Jul. 2019. arXiv: 1809.02701v4 [cs.CL]

I. Yamada, R. Tamaki, H. Shindo, and Y. Takefuji, "Studio Ousia's Quiz Bowl Question Answering System," *The NIPS '17 Competition: Building Intelligent Systems*, Sep. 2018. doi: 10.1007/978-3-319-94042-7_10

Z. Yang, P. Qing, S. Zhang, Y. Bengio, W. W. Cohen, R. Salakhutdinov, and C. D. Manning, "HotpotQA: A Dataset for Diverse, Explainable Multi-hop Question Answering," *Proceedings of the 2018 Conference on Empirical Methods in Natural Language Processing*, Sep. 2018. arXiv: 1809.09600v1 [cs.CL]

*Advice for Prospective Writers*, National Academic Quiz Tournaments. Accessed on: Sep. 24, 2021. [Online]. Available: https://www.naqt.com/jobs/application-advice.html

*High School Subject Distribution*, National Academic Quiz Tournaments. Accessed on: Oct. 11, 2021. [Online]. Available: https://www.naqt.com/hs/distribution.js

*Official NAQT Rules*, National Academic Quiz Tournaments. Accessed on: Sep. 24, 2021. [Online]. Available: https://www.naqt.com/rules/
17